\documentclass[conference]{IEEEtran}
\IEEEoverridecommandlockouts

\usepackage{cite}
\usepackage{amsmath,amssymb,amsfonts}
\usepackage{amsthm}
\usepackage{algorithmic}
\usepackage{graphicx}
\usepackage{textcomp}
\usepackage{xcolor}
\usepackage{array}
\usepackage{multirow}
\usepackage{booktabs}
\usepackage{float}
\usepackage{url}
\usepackage{enumitem}
\usepackage{hyperref}

\usepackage[most]{tcolorbox}

\newtcolorbox{caseInput}[1][]{%
    enhanced, breakable,
    colback=blue!4!white, colframe=blue!60!black,
    coltitle=white, fonttitle=\bfseries\small,
    boxrule=0.5pt, arc=2pt, left=4pt, right=4pt, top=2pt, bottom=2pt,
    title=Input: User Interaction History,
    #1}
\newtcolorbox{caseGT}[1][]{%
    enhanced, breakable,
    colback=green!4!white, colframe=green!55!black,
    coltitle=white, fonttitle=\bfseries\small,
    boxrule=0.5pt, arc=2pt, left=4pt, right=4pt, top=2pt, bottom=2pt,
    title=Ground Truth: Next Item,
    #1}
\newtcolorbox{caseDrift}[1][]{%
    enhanced, breakable,
    colback=orange!5!white, colframe=orange!75!black,
    coltitle=white, fonttitle=\bfseries\small,
    boxrule=0.5pt, arc=2pt, left=4pt, right=4pt, top=2pt, bottom=2pt,
    title=Drift Recovery: Vanilla NTP vs.\ HPR (beam $K{=}20$),
    #1}

\def\BibTeX{{\rm B\kern-.05em{\sc i\kern-.025em b}\kern-.08em
    T\kern-.1667em\lower.7ex\hbox{E}\kern-.125emX}}

\begin{document}
\pagestyle{plain}

\title{Bridging the Structural Gap: Adapting Autoregressive Generation for Recommendation}

\author{\IEEEauthorblockN{Junchao Zeng$^{\dagger}$}
\IEEEauthorblockA{\textit{Platform and Content Group} \\
\textit{Tencent}\\
Shenzhen, China\\
junchaozeng@tencent.com}
\and
\IEEEauthorblockN{Junzhang Zhu$^{\dagger}$}
\IEEEauthorblockA{\textit{Platform and Content Group} \\
\textit{Tencent}\\
Shenzhen, China\\
johannzhu@tencent.com}
\and
\IEEEauthorblockN{Junyang Chen}
\IEEEauthorblockA{\textit{CSSE} \\
\textit{Shenzhen University}\\
Shenzhen, China\\
junyangchen@szu.edu.cn}
\and
\IEEEauthorblockN{Yudong Li$^{*}$}
\IEEEauthorblockA{\textit{Platform and Content Group} \\
\textit{Tencent}\\
Shenzhen, China\\
elsonli@tencent.com}
\and
\IEEEauthorblockN{Wei Liu$^{*}$}
\IEEEauthorblockA{\textit{School of Artificial Intelligence} \\
\textit{Sun Yat-sen University}\\
Guangzhou, China\\
liuw259@mail.sysu.edu.cn}
\and
\IEEEauthorblockN{Chengxiang Zhuo}
\IEEEauthorblockA{\textit{Platform and Content Group} \\
\textit{Tencent}\\
Shenzhen, China}
\and
\IEEEauthorblockN{Zang Li}
\IEEEauthorblockA{\textit{Platform and Content Group} \\
\textit{Tencent}\\
Shenzhen, China}
\thanks{$^{\dagger}$These authors contributed equally to this work.}
\thanks{$^{*}$Corresponding authors.}
}

\maketitle

\begin{abstract}
Generative Recommendation (GR) has emerged as a new paradigm for sequential recommendation, in which a representative line of work encodes items into hierarchical semantic IDs via residual quantization and predicts the IDs token by token. However, this generative formulation still exhibits structural gaps with respect to the recommendation task: flattening multi-token IDs into a single sequence destroys item-level structure, and the inconsistency between training and inference over a hierarchical codebook gives rise to \emph{semantic drift}. To bridge these two gaps, we propose \textbf{BARGE}, which employs \textbf{Item Context-Aware Attention (ICA)} to restore item-level structure during encoding, and \textbf{Hierarchical Path Reranking (HPR)} together with \textbf{Dual-Path Decoding (DPD)} to suppress semantic drift from two complementary angles during decoding. Extensive experiments and analytical studies on public benchmarks and a large-scale offline test demonstrate that BARGE achieves superior recommendation performance. An online A/B test on a Tencent platform yields improvements of 0.60\% in click-through rate, 1.34\% in click unique visitors, and 1.70\% in total reading time, confirming the practical value of BARGE in industrial-scale recommendation.
\end{abstract}

\begin{IEEEkeywords}
Generative Recommendation, Semantics, Beam Search
\end{IEEEkeywords}

\section{Introduction}
\label{sec:intro}

\begin{figure*}[!t]
    \centering
    \includegraphics[width=\textwidth]{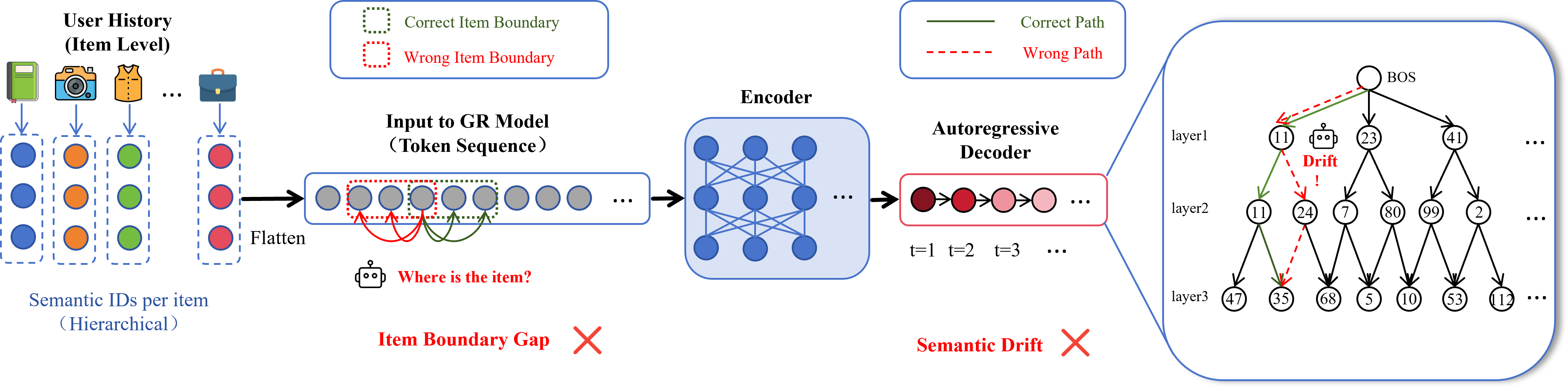}
    \caption{\small Two structural gaps in generative recommendation. \textbf{Left (Item-Boundary Gap, P1):} after each item is tokenized into $L$ hierarchical semantic IDs and flattened into a single sequence, item boundaries vanish from the input to the GR model, and the encoder can no longer tell which tokens belong to the same item (``Where is the item?''). \textbf{Right (Semantic Drift, P2):} on the hierarchical codebook tree, an error at any layer redirects decoding into a wrong subtree, so that the target leaf becomes unreachable along the chosen path no matter how the remaining tokens are scored.}
    \label{fig:three_problems}
\end{figure*}

Sequential recommendation aims to predict the next item that a user will interact with, given the chronologically ordered history of past interactions. The dominant approach has long been discriminative. Models learn to score every candidate item against the encoded user history and select the highest-scoring ones. However, the per-item scoring nature of this paradigm causes its storage and computation costs to scale linearly with the size of the item set, which fundamentally limits its scalability. Generative approaches sidestep this bottleneck by directly producing item identifiers autoregressively.

Generative Recommendation (GR)~\cite{de2020autoregressive,geng2022recommendation,rajput2023recommender,pang2025higr,hou2025actionpiece,guo2026onesug} offers a fundamentally different paradigm. Instead of scoring all candidates, GR generates the identifier of the target item token by token. To instantiate this generative idea, the mainstream practice~\cite{rajput2023recommender,wang2024learnable,yang2025sparse,lin2025order,liu2025generative,zheng2025universal,du2025vqrae} represents each item as a tuple of $L$ hierarchical semantic ID tokens $(c_1, c_2, \dots, c_L)$ obtained via residual quantization methods such as RQ-VAE~\cite{van2017neural,lee2022autoregressive}. This formulation allows semantically similar items to share ID prefixes and reduces the inference complexity from linear in the size of the item set to logarithmic in the depth of the codebook.

Despite this progress, the generative formulation was borrowed from natural language, where each token carries stand-alone lexical semantics. A hierarchical semantic codeword $c_l$, however, only acquires meaning when combined with its prefix $c_{<l}$. This mismatch manifests as two structural gaps:

\textbf{(P1) Loss of item-level structure.} Recommendation is fundamentally an item-level task, yet existing GR models decompose each item into $L$ semantic ID tokens and flatten all tokens into an undifferentiated sequence. The self-attention mechanism treats every token identically regardless of item membership, which effectively reduces item-level recommendation to token-level sequence modeling. As illustrated on the left of Fig.~\ref{fig:three_problems}, once items are flattened into a token sequence, the boundaries between items are no longer visible to the encoder, and tokens belonging to different items become indistinguishable from intra-item ones in terms of structural cues. Existing GR methods~\cite{rajput2023recommender,wang2024learnable,yang2025sparse} universally adopt this flat-sequence strategy and leave item boundaries to be recovered implicitly from positional signals alone.

\textbf{(P2) Semantic drift in hierarchical decoding.} Multi-level semantic IDs form a tree-like codebook in which the codeword selected at each layer constrains all subsequent layers, so that an error at any layer redirects the search into a wrong subtree. We refer to this family of failure modes collectively as \emph{semantic drift}. As illustrated on the right of Fig.~\ref{fig:three_problems}, once decoding deviates from the correct branch, the target leaf is no longer reachable along the chosen path, while standard beam search keeps selecting codewords by locally normalized probabilities and is unaware of such cumulative drift. On TIGER, the per-layer accuracy of $c_3$ drops from 77.0\% (correct prefix) to 0.6\% (erroneous prefix), a 128$\times$ gap~(Section~\ref{sec:motivation}). This drift can be attacked from two orthogonal angles: intra-path, by reranking candidates within a single channel via global path-level coherence; and cross-channel, by exposing the decoder to a second, structurally orthogonal quantization channel so that an item missed by one channel can still be recovered through the other via OR-style fusion.

To bridge these two gaps, we propose \textbf{BARGE} (\textbf{B}ridging \textbf{A}uto\textbf{R}egressive \textbf{G}eneration for r\textbf{E}commendation), with three lightweight modules: \textbf{ICA} closes the encoder-side gap (P1), while \textbf{HPR} and \textbf{DPD} jointly close the decoder-side drift gap (P2) from the two complementary angles above. The main contributions of this work are summarized as follows:

\begin{itemize}
    \item We formalise the long-standing semantic-fidelity gap of GR as two structural gaps along the GR pipeline: an encoder-side item-boundary gap, and a decoder-side semantic-drift gap. We further show that the latter can be mitigated from two complementary, orthogonal angles.
    \item We propose \textbf{BARGE} with three lightweight and mutually orthogonal modules. \textbf{ICA} restores item-level structure at the encoder; \textbf{HPR} suppresses semantic drift via per-layer dual-tower contrastive reranking within a single decoding channel; and \textbf{DPD} further mitigates the same drift from a complementary angle via a Dual-Decoder coupled by an OSQ-VAE tokenizer.
    \item Extensive experiments demonstrate consistent improvements over strong baselines. BARGE has also been deployed on Tencent's commercial media platform, bringing consistent improvements across core engagement metrics in real-world industrial scenarios.
\end{itemize}

\section{Related Work}

\subsection{Discriminative and Generative Recommendation}

Mainstream sequential recommendation methods are discriminative: they encode the interaction history into a user representation and score candidate items, with architectures evolving from RNN-based GRU4Rec~\cite{hidasi2015session} and CNN-based Caser~\cite{tang2018caser} to Transformer-based SASRec~\cite{kang2018self}, BERT4Rec~\cite{sun2019bert4rec}, FDSA~\cite{zhang2019feature}, S$^3$-Rec~\cite{zhou2020s3}, and more recent variants based on diffusion~\cite{li2023diffurec}, sparse attention~\cite{li2023strec}, and zero-shot transfer~\cite{chen2026zero}. Despite their architectural diversity, these methods all rely on a candidate-scoring paradigm.

Generative Recommendation (GR)~\cite{cheng2016wide,covington2016deep,guo2017deepfm,qiu2025one} instead formulates recommendation as autoregressive generation of discrete item identifiers~\cite{de2020autoregressive,geng2022recommendation,cui2022m6rec,li2024survey,bao2025bi,zhang2025reinforced}. TIGER~\cite{rajput2023recommender} established a representative framework that encodes items into hierarchical semantic IDs via RQ-VAE~\cite{lee2022autoregressive}, and subsequent work improves upon it through collaborative signals (LETTER~\cite{wang2024learnable}), hybrid representations (COBRA~\cite{yang2025sparse}), knowledge integration (MVIGER~\cite{kim2024mviger}), and alternative indexing schemes~\cite{hua2023index,sun2023learning}. More recently, HSTU~\cite{zhai2024actions}, OneRec~\cite{deng2025onerec,liu2025onerec}, RPG~\cite{hou2025generating}, DACT~\cite{feng2026drift}, DELRec~\cite{zhang2025delrec}, and Reg4Rec~\cite{xing2025reg4rec} further advance GR through scalable architectures, end-to-end modeling, efficient generation, continual tokenization, knowledge distillation, and reasoning enhancement, respectively. Unlike Reg4Rec, the proposed DPD explicitly derives two complementary channels from a learnable orthogonal rotation of encoder features, providing a structural guarantee of channel diversity.

\subsection{Structural Awareness and Semantic Drift Suppression}

\textit{Structural awareness in the encoding stage.}
When each item is decomposed into $L$ semantic ID tokens, the self-attention in the encoder treats all tokens as an undifferentiated flat sequence and consequently loses item-level grouping. In NLP, ALiBi~\cite{press2021train} and Longformer~\cite{beltagy2020longformer} introduce structured attention via relative position bias and local-global patterns, but neither is tailored to semantic IDs. TrieRec~\cite{xu2026trie} explores leveraging the prefix tree structure to introduce structural bias. Orthogonal to these efforts, the proposed ICA enriches every token with item-level semantics prior to encoding, via cross-attention pooling followed by a gated residual injection.

\textit{Semantic drift suppression in hierarchical codebook decoding.}
Hierarchical semantic drift is inherent to any GR method that decodes multi-level semantic IDs, which distinguishes it from exposure bias in flat-vocabulary generation. Existing methods apply standard beam search without consistency-aware modification. PROMISE~\cite{guo2026promise} explores applying the process reward idea from LLMs to rerank step-by-step decoding. APAO~\cite{yu2026apao} addresses the gap between training and inference by introducing prefix-level pointwise and pairwise ranking losses together with an adaptive worst-prefix weighting strategy that aligns training with beam-search inference. In contrast, the proposed HPR is a lightweight, label-free auxiliary. At each decoder layer, it scores the semantic compatibility between the history-aggregated hidden state and the cumulative path embeddings via a dual-tower contrastive objective trained with symmetric InfoNCE.

\section{Preliminary}

Let the user set be $\mathcal{U}$ and the item set be $\mathcal{V}$. For a user $u \in \mathcal{U}$, the chronologically ordered historical interaction sequence is denoted as $\mathbf{s}_u = (v_1, v_2, \dots, v_T)$, where $v_t \in \mathcal{V}$. The goal is to predict the next item $v_{T+1}$.

\textit{Semantic IDs via RQ-VAE.}
In the generative recommendation framework, each item $v$ is assigned a discrete semantic ID of length $L$ via Residual-Quantized VAE (RQ-VAE)~\cite{van2017neural,lee2022autoregressive}, which iteratively quantizes the residual of a pretrained item embedding across $L$ codebooks $\{\mathcal{C}_l\}_{l=1}^{L}$:
\begin{equation}
v \;\rightarrow\; \mathbf{s}^{(v)} = (c_1^{(v)},\, c_2^{(v)},\, \dots,\, c_L^{(v)}),
\end{equation}
where $c_l^{(v)} \in \mathcal{C}_l$ is the codeword assigned at layer $l$. Shallower layers capture coarse-grained semantic categories, whereas deeper layers progressively refine the representation within narrower subtrees. The codebook configuration and training details are discussed in Section~\ref{sec:codebook}.

\textit{Generative recommendation.}
The user history is represented as a flattened token sequence $\mathbf{X} = [\mathbf{s}^{(v_1)}; \cdots; \mathbf{s}^{(v_T)}] \in \mathbb{R}^{(T \cdot L) \times d}$, which is fed into an encoder~\cite{vaswani2017attention} that produces the encoded history $\mathbf{H} \in \mathbb{R}^{(T \cdot L) \times d}$ and autoregressively generates the target semantic ID of the next item:
\begin{equation}
P(v_{T+1} \mid \mathbf{s}_u) = \prod_{l=1}^{L} P_l\bigl(c_l \mid c_{<l},\, \mathbf{H}\bigr),
\label{eq:standard_ar}
\end{equation}
where $\mathbf{H}$ is the encoder output and $P_l$ is the decoder distribution over codebook $\mathcal{C}_l$ at decoding layer $l$.

\section{Methodology}

\subsection{Model Overview}

\begin{figure*}[t]
    \centering
    \includegraphics[width=\textwidth]{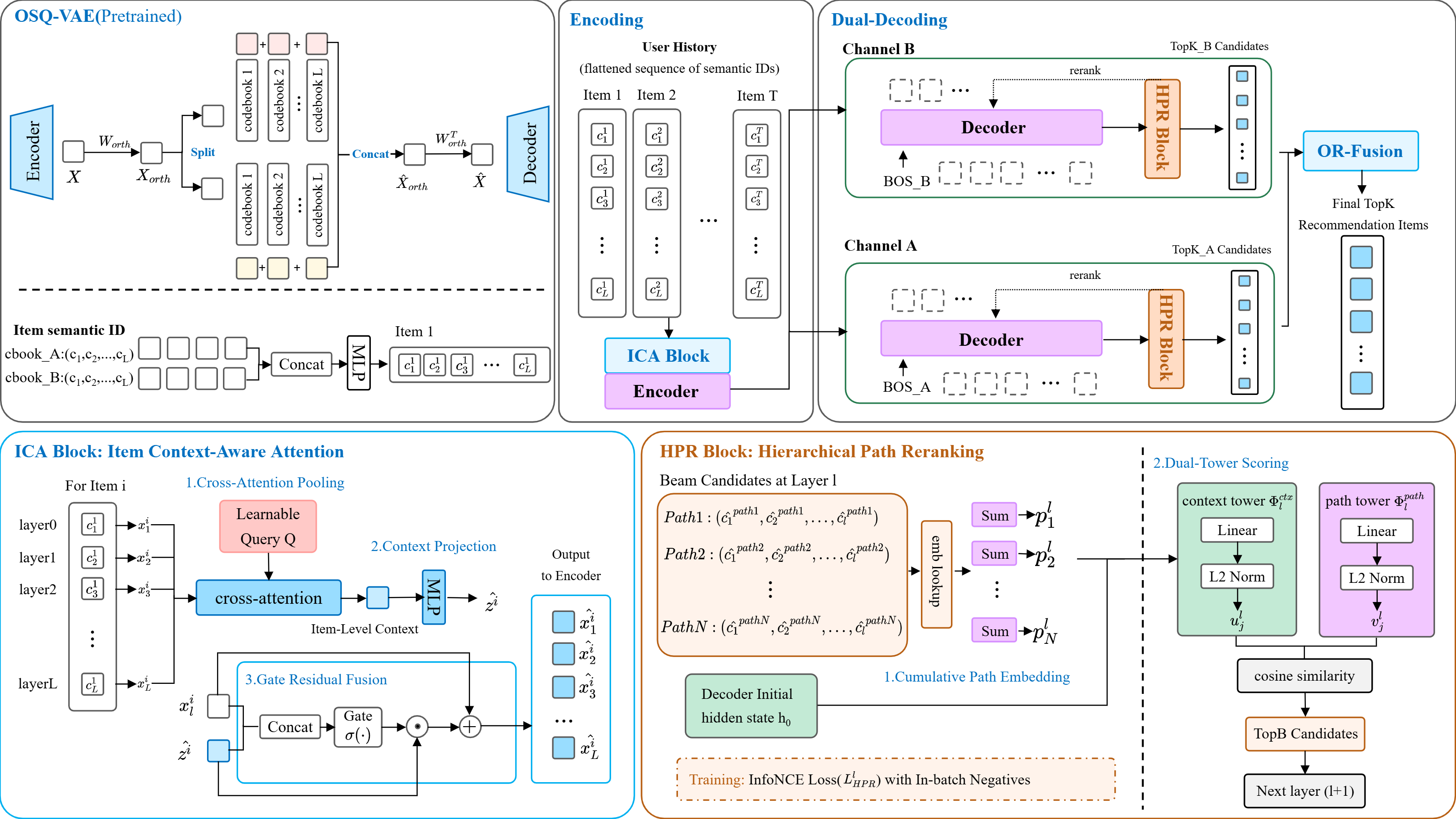}
    \caption{\small Overview of BARGE. (Top-left) The pretrained OSQ-VAE tokenizes each item into two channel-specific semantic-ID tuples (cbook\_A and cbook\_B) via orthogonal rotation and dual codebook stacks; the dual IDs are projected by an MLP into the flattened token sequence. (Top-middle) The user history passes through the ICA before the shared encoder. (Top-right) Two independent decoders (Channel A and Channel B) each followed by an HPR run in parallel; their Top-$K$ candidate lists are merged by OR-Fusion to produce the final recommendations. (Bottom-left) ICA: cross-attention pooling with a learnable query, MLP projection, and gated residual fusion. (Bottom-right) HPR: per-layer dual-tower scoring of cumulative path embeddings against the decoder hidden state $h_0$, trained with $\mathcal{L}_{\mathrm{HPR}}$.}
    \label{fig:overview}
\end{figure*}

As illustrated in Fig.~\ref{fig:overview}, BARGE addresses the two structural gaps identified in Section~\ref{sec:intro} with three modules: one that targets the encoder-side gap, and two complementary modules that jointly close the decoder-side drift gap. \textbf{(1) Item Context-Aware Attention (ICA)} aggregates all token embeddings within each item into a single item-level context $\mathbf{z}^{(i)}$ and fuses it back into every token via a gating network before the encoder block, which closes the item-boundary gap (P1). \textbf{(2) Hierarchical Path Reranking (HPR)} augments beam search with per-layer dual-tower contrastive scoring trained via the symmetric InfoNCE loss ($\mathcal{L}_{\text{HPR}}$) alongside the standard next-token prediction loss ($\mathcal{L}_{\text{NTP}}$), in order to suppress the intra-path facet of decoder-side drift along a single quantization channel. \textbf{(3) Dual-Path Decoding (DPD)} replaces the single RQ-VAE tokenizer with an OSQ-VAE tokenizer. The OSQ-VAE decomposes the encoder feature into two complementary subspaces via a learnable orthogonal rotation $R$, and emits two semantic-ID tuples per item. On top of a shared encoder, DPD runs a Dual-Decoder with channel-private HPR, and fuses their per-channel candidate items in the item-id space at inference time. In this way, the cross-channel facet of decoder-side drift is suppressed through orthogonal quantization channels.

\subsection{Structure-Aware Encoding via Item Context-Aware Attention (ICA)}
\label{sec:ica}

ICA adopts an aggregate-then-fuse strategy (Fig.~\ref{fig:overview}, bottom-left): for each item, it first aggregates all token embeddings into a single item-level context, and then fuses this context back into every token via a gating network.

\textit{Cross-attention pooling.}
Given the $L$ token embeddings $\{\mathbf{x}_1^{(i)}, \dots, \mathbf{x}_L^{(i)}\}$ of item $v_i$, ICA uses a learnable query vector $\mathbf{q} \in \mathbb{R}^d$ to compute the item-level context representation via cross-attention, in which $\mathbf{q}$ serves as the query and the tokens of the item serve as the keys and values:
\begin{equation}
\mathbf{z}^{(i)} = \text{LayerNorm}\left(\text{CrossAttn}(\mathbf{q},\; \mathbf{X}^{(i)},\; \mathbf{X}^{(i)})\right),
\end{equation}
where $\mathbf{X}^{(i)} = [\mathbf{x}_1^{(i)}; \dots; \mathbf{x}_L^{(i)}] \in \mathbb{R}^{L \times d}$ is the matrix of $L$ token embeddings for item $v_i$, and $\mathbf{z}^{(i)} \in \mathbb{R}^d$ is the resulting item-level representation. Cross-attention is used here so that the learnable query $\mathbf{q}$ can adaptively and dynamically weight the contribution of each token when forming the item-level context, since different layers of a semantic ID encode relative information at different granularities.

\textit{Context projection.}
The context vector undergoes multi-layer nonlinear transformation:
\begin{equation}
\hat{\mathbf{z}}^{(i)} = W_2 \cdot \text{GELU}(W_1 \cdot \mathbf{z}^{(i)} + \mathbf{b}_1) + \mathbf{b}_2,
\end{equation}
where $W_1 \in \mathbb{R}^{d_f \times d}$, $W_2 \in \mathbb{R}^{d \times d_f}$, $d_f$ is the feed-forward hidden dimension, and $\hat{\mathbf{z}}^{(i)} \in \mathbb{R}^d$ is the projected context vector after the nonlinear transformation.

\textit{Gated residual fusion.}
For each token $l \in \{1, \dots, L\}$ of item $v_i$, a gating network controls how much item-level context is fused back into each token:
\begin{equation}
\mathbf{g}_l^{(i)} = \sigma\left(W_g \cdot [\mathbf{x}_l^{(i)} \| \hat{\mathbf{z}}^{(i)}] + \mathbf{b}_g\right),
\end{equation}
\begin{equation}
\hat{\mathbf{x}}_l^{(i)} = \mathbf{x}_l^{(i)} + \mathbf{g}_l^{(i)} \odot \hat{\mathbf{z}}^{(i)},
\end{equation}
where $W_g \in \mathbb{R}^{d \times 2d}$, $\mathbf{g}_l^{(i)} \in \mathbb{R}^d$ is the gate vector, $\sigma(\cdot)$ is the sigmoid function, $\|$ denotes concatenation, and $\odot$ denotes element-wise multiplication. The gated residual form allows each token to decide how much item-level context to admit, so that the injected context never overwhelms the positional and layer-specific information already encoded in $\mathbf{x}_l^{(i)}$. When the gate approaches zero, ICA reduces to the identity mapping and preserves the original token representation; when the gate approaches one, the token is fully merged with the item context. We empirically observe that the learned gate stabilises around $0.35$--$0.38$ across all four layers (Section~\ref{sec:qualitative}), which confirms that the network indeed learns a moderate and layer-consistent fusion strength.

\subsection{Hierarchical Path Reranking (HPR)}
\label{sec:hpr}

HPR augments beam search with a per-layer reranking mechanism that evaluates the semantic compatibility between the initial hidden state of the decoder $\mathbf{h}_0$ and the cumulative path embedding at each decoding layer (Fig.~\ref{fig:overview}, bottom-right). The key insight is that $\mathbf{h}_0$, produced by the cross-attention of the decoder over the full encoder output before any token is generated, serves as a holistic representation of the historical preferences of the user, which makes it a natural and suitable anchor for assessing whether a candidate path aligns with the intent of the user.

\textit{Cumulative path embedding.}
Unlike token-level scoring that evaluates individual codewords in isolation, HPR operates explicitly at the path level to better capture dependencies between different layers. For each candidate at layer $l$, the cumulative path embedding is computed as
\begin{equation}
\mathbf{p}^{(l)} = \sum_{j=1}^{l} \mathbf{e}_{c_j},
\end{equation}
where $\mathbf{e}_{c_j} \in \mathbb{R}^{d_{\text{emb}}}$ is the learnable embedding of codeword $c_j$. This cumulative representation captures the semantic trajectory of the partial path and enables the reranker to assess global coherence rather than evaluating each token independently.

\textit{Per-layer dual-tower scoring.}
As illustrated in Fig.~\ref{fig:overview} (bottom-right), HPR maintains a separate dual-tower scorer for each layer $l$. Each scorer projects the initial hidden state of the decoder and the cumulative path embedding into a shared low-dimensional space, and then computes their cosine similarity scaled by a learnable temperature:
\begin{equation}
r_l(\mathbf{h}_0, \mathbf{p}^{(l)}) = \cos\!\bigl(\phi_l^{\text{ctx}}(\mathbf{h}_0),\; \phi_l^{\text{path}}(\mathbf{p}^{(l)})\bigr) \cdot e^{\tau_l},
\end{equation}
where $\phi_l^{\text{ctx}}: \mathbb{R}^{d_{\text{attn}}} \to \mathbb{R}^{d_{\text{proj}}}$ and $\phi_l^{\text{path}}: \mathbb{R}^{d_{\text{emb}}} \to \mathbb{R}^{d_{\text{proj}}}$ are layer-specific linear projections followed by $L_2$ normalisation, and $\tau_l$ is the learnable log-temperature. The dual-tower architecture allows efficient scoring of multiple candidate paths, as the context projection is computed only once per sample.

\textit{Training via symmetric InfoNCE.}
HPR is jointly trained with the main model using the InfoNCE loss. For each layer $l$, within a batch of $B$ samples, the positive pair consists of the initial hidden state of the decoder $\mathbf{h}_0^{(i)}$ and the ground-truth cumulative path embedding $\mathbf{p}^{(l,i)}$ for sample $i$. In addition to the in-batch negatives, we incorporate prefix-aware negatives that explicitly expose the model to plausible but erroneous prefixes. Concretely, at each decoding step, we draw high-probability candidates that are not ground truth from the NTP distribution, and further introduce business-level negatives such as impressed but unclicked items. These prefix-conditioned negatives emulate the drift patterns encountered during inference and strengthen the model's ability to discriminate correct semantic prefixes from misleading or user-irrelevant ones. The per-layer loss is defined as
\begin{equation}
\mathcal{L}_{\text{HPR}}^{(l)} = \frac{1}{2}\left[\mathcal{L}_{\text{c2p}}^{(l)} + \mathcal{L}_{\text{p2c}}^{(l)}\right],
\end{equation}
where $\mathcal{L}_{\text{c2p}}^{(l)}$ (context-to-path) and $\mathcal{L}_{\text{p2c}}^{(l)}$ (path-to-context) are standard cross-entropy losses over the similarity matrix. Taking $\mathcal{L}_{\text{c2p}}^{(l)}$ as an example,
\begin{equation}
\mathcal{L}_{\text{c2p}}^{(l)} \!=\! -\frac{1}{B}\!\sum_{i=1}^{B} \log \frac{\exp\bigl(r_l^{(i,i)}\bigr)}{\sum_{j=1}^{B} \exp\bigl(r_l^{(i,j)}\bigr)},
\end{equation}
where $r_l^{(i,j)} = r_l(\mathbf{h}_0^{(i)}, \mathbf{p}^{(l,j)})$ denotes the reranker score between the $i$-th context and the $j$-th path.
The symmetric formulation ensures that both projection heads are well-calibrated: the context tower learns to produce representations that are discriminative across different paths, while the path tower learns representations that are discriminative across different user contexts. The total reranker loss is averaged across all $L$ layers as $\mathcal{L}_{\text{HPR}} = \frac{1}{L}\sum_{l=1}^{L} \mathcal{L}_{\text{HPR}}^{(l)}$.

\textit{Joint scoring during inference.}
Standard beam search keeps a set of $B$ partial paths and, at layer $l$, expands them into $B\times|\mathcal{C}_l|$ candidates scored solely by their cumulative log-probability; the top-$B$ are kept for the next layer. Such greedy local selection ignores global semantic coherence, and simply enlarging $B$ is costly, since the KV cache and the expanded score tensor all scale linearly with $B$. We therefore follow prior generative works (e.g., $B=20$ in TIGER~\cite{rajput2023recommender}) and instead correct erroneous paths without enlarging the beam.

HPR intervenes after beam expansion at each layer. From the $B\times|\mathcal{C}_l|$ expanded candidates, we first keep the top-$N$ (with $B<N\ll B\times|\mathcal{C}_l|$) ranked by generation log-probability to form a scoring pool, which retains semantically reasonable but locally lower-probability paths that vanilla beam search would prematurely discard. Subsequently, each candidate in the pool is rescored by fusing its original generation log-probability with the score from the HPR reranker.
\begin{equation}
\begin{split}
\text{score}(c, l) = \; & \log p(c \mid c_{<l}, \mathbf{s}_u) \\
& + \lambda \cdot \log\text{softmax}\!\left(r_l(\mathbf{h}_0, \mathbf{p}^{(l)})\right),
\end{split}
\end{equation}
where $c$ is a candidate codeword at layer $l$, $\mathbf{p}^{(l)}$ is the cumulative path embedding obtained by appending $c$ to $c_{<l}$, $p(c \mid c_{<l}, \mathbf{s}_u)$ is the softmax generation probability and $\lambda \ge 0$ controls the reranking strength ($\lambda = 0$ recovers vanilla beam search). The top-$B$ paths under this fused score advance to layer $l{+}1$, so the outgoing beam width is unchanged and only a lightweight dual-tower scoring is added per layer, while the enlarged pool $N$ allows semantically inconsistent paths to be corrected before their errors propagate deeper.

\subsection{Dual-Path Decoding (DPD)}
\label{sec:dpd}

ICA and HPR both operate within a single quantization channel, and neither addresses the cross-channel facet. A single RQ-VAE structure commits the decoder to one factorization of an item, projecting its rich semantics onto a single quantization axis; any semantic facet of the item not captured by that axis is permanently locked outside the corresponding semantic subtree. DPD tries to close this gap with three components: an OSQ-VAE tokenizer, a Dual-Decoder, and an OR-fusion inference procedure.

\textit{OSQ-VAE (Orthogonal Split-and-Quantize VAE) tokenizer.}
Let $\mathbf{z}\!\in\!\mathbb{R}^{D}$ be the pretrained item embedding. The OSQ-VAE tokenizer first applies a learnable orthogonal rotation
\begin{equation}
\tilde{\mathbf{z}} \;=\; R\,\mathbf{z}, \qquad R \in \mathbb{R}^{D\times D},\;\; R^{\top}R = I_D,
\label{eq:dpd_rot}
\end{equation}
in which $R$ is parameterized as a product of Householder reflections, so that $R^{\top}R = I_D$ holds by construction throughout training without any auxiliary loss term. The rotated feature is then split into two equal halves
\begin{equation}
\tilde{\mathbf{z}} \;=\; \bigl[\,\tilde{\mathbf{z}}^{(A)} \,\|\, \tilde{\mathbf{z}}^{(B)}\,\bigr],\quad \tilde{\mathbf{z}}^{(A)},\tilde{\mathbf{z}}^{(B)} \!\in\! \mathbb{R}^{D/2},
\label{eq:dpd_split}
\end{equation}
and each half is quantized by an independent $L$-layer residual codebook stack, which yields per-channel semantic-ID tuples $\mathbf{s}^{(c)}\!=\!(c_1^{(c)},\dots,c_L^{(c)})$ for $c\!\in\!\{A,B\}$. Because $R^{\top}R\!=\!I_D$ and the split is coordinate-aligned in the rotated frame, the supporting subspaces of the two channels satisfy $S_A\!\perp\! S_B$ and $S_A\!\oplus\! S_B\!=\!\mathbb{R}^{D}$, which is a hard architectural invariant. The OSQ-VAE is trained end-to-end with the standard reconstruction and commitment losses applied independently to each channel:
\begin{equation}
\begin{aligned}
\mathcal{L}_{\mathrm{OSQ}} \;=\;& \underbrace{\|\mathbf{z} - \hat{\mathbf{z}}\|_2^2}_{\mathcal{L}_{\mathrm{recon}}} + \sum_{c\in\{A,B\}} \underbrace{\|\mathrm{sg}[\tilde{\mathbf{z}}^{(c)}] - \hat{\tilde{\mathbf{z}}}^{(c)}\|_2^2}_{\text{codebook loss}} \\
&+\; \beta \sum_{c\in\{A,B\}} \underbrace{\|\tilde{\mathbf{z}}^{(c)} - \mathrm{sg}[\hat{\tilde{\mathbf{z}}}^{(c)}]\|_2^2}_{\text{commitment loss}},
\end{aligned}
\label{eq:dpd_osq}
\end{equation}
where $\hat{\mathbf{z}}\!=\!R^{\top}[\hat{\tilde{\mathbf{z}}}^{(A)}\!\|\hat{\tilde{\mathbf{z}}}^{(B)}]$ is the rotated-back reconstruction, $\mathrm{sg}[\cdot]$ denotes the stop-gradient operator, and $\beta$ is the commitment weight. The first term $\mathcal{L}_{\mathrm{recon}}$ is the global reconstruction loss between the encoder output $\mathbf{z}$ and the dequantized reconstruction $\hat{\mathbf{z}}$. Inside the summation, the two per-channel terms differ only in the direction of stop-gradient: the codebook loss freezes the encoder output $\tilde{\mathbf{z}}^{(c)}$ and pulls the codebook embedding $\hat{\tilde{\mathbf{z}}}^{(c)}$ towards it, while the commitment loss freezes the codebook embedding and forces the encoder output to commit to it, weighted by $\beta$.

\textit{Dual-Decoder.}
Given a user history, BARGE encodes it once through the shared ICA-enriched encoder. Two decoder towers, $\mathrm{Dec}^{(A)}$ and $\mathrm{Dec}^{(B)}$, then run in parallel on top of the shared encoder output. Each tower has its own input projection, per-layer output heads tied to its channel-specific codebooks, and its own HPR scorer with channel-private projection heads $\bigl(\phi_l^{\mathrm{ctx},(c)},\,\phi_l^{\mathrm{path},(c)}\bigr)$. DPD follows a two-stage training procedure. In the first stage, the OSQ-VAE is pre-trained offline via $\mathcal{L}_{\mathrm{OSQ}}$ (Eq.~\eqref{eq:dpd_osq}) to produce two sets of channel-specific semantic IDs $\{\mathbf{s}^{(A)}\}$ and $\{\mathbf{s}^{(B)}\}$ for all items, after which the weights of the OSQ-VAE are frozen. In the second stage, the Dual-Decoder is trained on top of the fixed semantic IDs. Decoder $\mathrm{Dec}^{(c)}$ is trained to autoregressively predict the channel-$c$ semantic ID:
\begin{equation}
P\bigl(v_{T+1}^{(c)} \mid \mathbf{s}_u\bigr) \;=\; \prod_{l=1}^{L} P_l^{(c)}\!\bigl(c_l^{(c)} \,\bigm|\, c_{<l}^{(c)},\, \mathbf{H}\bigr).
\end{equation}
The overall training loss sums the next-token prediction loss and the HPR contrastive loss over both channels:
\begin{equation}
\mathcal{L}_{\text{total}} \;=\; \sum_{c\in\{A,B\}} \Bigl(\mathcal{L}_{\mathrm{NTP}}^{(c)} + \mathcal{L}_{\mathrm{HPR}}^{(c)}\Bigr).
\label{eq:total_loss}
\end{equation}
Because the two towers do not share the decoder parameters or the HPR scorers, each tower specializes in the codebook statistics of its own channel, while the shared encoder benefits from gradients flowing in from both sides.

\textit{Inference: OR-fusion in the item-id space.}
\label{sec:dpd_fusion}
At inference time, each tower runs an independent beam search of width $B$ and returns a channel-specific top-$K$ item list $A_K$ and $B_K$ over the shared item-id space (via the two OSQ-VAE codebooks). We form the union pool $U \triangleq A_K \cup B_K$ (so $K \le |U| \le 2K$) and rescore every $v\in U$ with an OR-fusion operator:
\begin{equation}
s(v) \;=\; f\!\bigl(s^{(A)}(v),\,s^{(B)}(v)\bigr),
\label{eq:dpd_fusion}
\end{equation}
where items appearing in only one channel take a sentinel low value on the absent side. The final ranked list is obtained by sorting $U$ by $s(v)$ and \emph{truncating back to top-$K$}, on which Recall@$K$ and NDCG@$K$ are computed. This OR semantics recovers any item that at least one channel ranks highly \emph{and} that survives the fusion truncation, and its effectiveness relies on the two channels proposing complementary top-$K$ candidates rather than on channel independence. Since each tower still uses the same beam width $B$ and the merged list is truncated at the same $K$ as single-tower baselines, all metrics are evaluated under exactly the same protocol. We validate the complementarity together with the choice of $f$ in Section~\ref{sec:ablation_dpd}.

\subsection{Design Rationale and Verifiable Conditions}
\label{sec:theory}

We complement the design of BARGE with a module-wise analysis. For each module, we derive a condition that characterizes when it improves recommendation accuracy, and link that condition to the empirical section in which it is directly measured.

\textit{ICA: identity-preserving property.}
Since $\hat{\mathbf{x}}_l^{(i)} = \mathbf{x}_l^{(i)} + \mathbf{g}_l^{(i)} \odot \hat{\mathbf{z}}^{(i)}$ with $\mathbf{g}_l^{(i)} \in [0,1]^d$, we have $\|\hat{\mathbf{x}}_l^{(i)} - \mathbf{x}_l^{(i)}\|_2 \le \|\hat{\mathbf{z}}^{(i)}\|_2$, and the gate can shrink to $\mathbf{0}$ to recover the vanilla encoder. ICA therefore augments rather than overwrites the original signal.

\textit{HPR: decomposing per-layer errors.}
At layer $l$, let $g_l$ be the autoregressive likelihood, $r_l$ the HPR reranker score, and $\pi_l \propto g_l\cdot r_l^{\lambda}$ the fused score used by HPR. Denote the per-layer miss events of vanilla beam search and of HPR by
$\varepsilon_l^{\mathrm{van}} \!=\! \Pr[c_l^* \!\notin\! \mathrm{Top}_B(g_l)]$ and
$\varepsilon_l^{\mathrm{HPR}} \!=\! \Pr[c_l^* \!\notin\! \mathrm{Top}_B(\pi_l)]$, respectively. Reranking with $\pi_l$ moves the boundary of the surviving top-$B$ in two directions: it can pull a previously discarded ground-truth codeword into the top-$B$ (\textit{rescue}), but it can also push a previously surviving ground-truth codeword out of it (\textit{damage}). Formally, define
\begin{align*}
\mathrm{Rescue}_l &\;\triangleq\; \{\,c_l^* \in \mathrm{Top}_B(\pi_l)\setminus \mathrm{Top}_B(g_l)\,\},\\
\mathrm{Damage}_l &\;\triangleq\; \{\,c_l^* \in \mathrm{Top}_B(g_l)\setminus \mathrm{Top}_B(\pi_l)\,\}.
\end{align*}
Then the following identity holds without any assumption on $r_l$:
\begin{equation}
\varepsilon_l^{\mathrm{van}} - \varepsilon_l^{\mathrm{HPR}} \;=\; \Pr[\mathrm{Rescue}_l] - \Pr[\mathrm{Damage}_l].
\label{eq:hpr_rescue_damage}
\end{equation}
Eq.~\eqref{eq:hpr_rescue_damage} converts the question ``when does HPR help?'' into a directly measurable one: HPR is net beneficial at layer $l$ iff $\Pr[\mathrm{Rescue}_l]>\Pr[\mathrm{Damage}_l]$. We do not assume this inequality a priori. Two pieces of evidence already in the paper are consistent with it. First, the InfoNCE training of $r_l$ maximises a lower bound on the mutual information between the user context and the cumulative path~\cite{oord2018representation}, which biases $\mathrm{Rescue}_l$ above $\mathrm{Damage}_l$ for ground-truth-aligned paths. Second, the inverted-U behavior of $\lambda$ in Section~\ref{sec:ablation_dpd} (Fig.~\ref{fig:hpr_sensitivity}) is the predicted shape under Eq.~\eqref{eq:hpr_rescue_damage}: at small $\lambda$ the reranker barely perturbs the boundary, so $\Pr[\mathrm{Damage}_l]\!\approx\!0$ and any non-zero rescue translates into a gain; at overlarge $\lambda$ the reranker overrides the likelihood and inflates $\Pr[\mathrm{Damage}_l]$, eventually erasing the gain.

\textit{DPD: an OR-fusion gain identity at the union-pool level.}
Let $E^{(A)}$, $E^{(B)}$ be the per-channel top-$K$ miss events and $E^{U}\triangleq E^{(A)}\cap E^{(B)}$ the event that the ground truth is absent from $U=A_K\cup B_K$. By construction, $E^{U}$ occurs iff both channels miss, which is a set-theoretic identity independent of $f$. With $\kappa\triangleq\Pr[E^{(B)}\mid E^{(A)}]$, the chain rule yields
\begin{equation}
\underbrace{\Pr[E^{(A)}] - \Pr[E^{U}]}_{\text{union-pool miss reduction over channel A}} \;=\; (1-\kappa)\cdot\Pr[E^{(A)}].
\label{eq:dpd_or_gain}
\end{equation}
Eq.~\eqref{eq:dpd_or_gain} lives at the union-pool level and therefore upper-bounds the top-$K$ miss reduction achievable by any $f$ under the fixed-$K$ protocol, since rescoring $U$ and truncating back to $K$ may still push an item present in $U$ out of the final top-$K$. It thus gives a necessary condition $\kappa\!<\!1$, i.e., non-trivial channel complementarity, and turns DPD's design question into a measurable one: how small is $\kappa$? The orthogonal rotation $R$ in OSQ-VAE ($R^\top R=I$, $S_A\perp S_B$) is what drives it down. We verify $\kappa\!<\!1$ in Section~\ref{sec:ablation_dpd}: the Jaccard overlap between the two top-$K$ pools is only $0.18$/$0.17$ on Beauty/Sports and $15$--$24\%$ of union-pool hits come from a single channel; how much of this union-pool headroom is realized as fixed-$K$ gain is then determined by $f$ (Section~\ref{sec:ablation_dpd}) and reflected in Table~\ref{tab:overall_amazon}.

\textit{Stacking along orthogonal failure dimensions.}
Eq.~\eqref{eq:hpr_rescue_damage} and Eq.~\eqref{eq:dpd_or_gain} act on disjoint failure modes: HPR rescues a ground-truth codeword that is still reachable within a single channel, whereas DPD recovers an item that is no longer reachable in one channel by exposing it through the other. Because the two mechanisms reduce different terms of the overall miss probability, we expect their gains to be largely additive, which is consistent with the component-wise ablation in Section~\ref{sec:ablation}.

\section{Experiment}

\subsection{Experimental Setup}

\textit{Datasets.}
We evaluate the proposed method on three datasets: (1) Amazon Beauty and (2) Amazon Sports and Outdoors, two widely used sequential recommendation benchmarks with 5-core filtering~\cite{he2016ups}; and (3) a large-scale offline test from Tencent's commercial media platform that comprises hundreds of millions of users and interactions over 11 days (the first 10 days are used for training and the last day for evaluation). Statistics of the Amazon datasets are summarized in Table~\ref{tab:dataset}. For the Amazon datasets, we follow the preprocessing of P5~\cite{geng2022recommendation} and adopt the leave-one-out evaluation protocol.

\begin{table}[H]
\centering
\caption{Statistics of the datasets.}
\label{tab:dataset}
\begin{tabular}{p{2.0cm} >{\centering}p{1.5cm} >{\centering}p{1.3cm} >{\centering\arraybackslash}p{1.78cm}}
\toprule
Dataset & \#Users & \#Items & \#Interactions \\
\midrule
Amazon Beauty & 22,363 & 12,101 & 198,502 \\
Amazon Sports & 25,598 & 18,357 & 296,337 \\
Industry      & $>$100M & $>$300k & $>$100M \\
\bottomrule
\end{tabular}
\end{table}

\textit{Metrics.}
All datasets are evaluated with Recall@K and NDCG@K, with $K \in \{5, 10\}$. All metrics are computed on the full item set.

\textit{Baselines.}
We compare the proposed method against two groups of competitors.

\noindent \textit{Non-generative sequential recommenders:}
\begin{itemize}
    \item \textbf{P5}~\cite{geng2022recommendation}: reformulates recommendation tasks as text-to-text generation under a shared language model.
    \item \textbf{Caser}~\cite{tang2018caser}: treats the recent interaction history as an image and applies horizontal and vertical convolutions to capture sequential patterns.
    \item \textbf{HGN}~\cite{ma2019hierarchical}: models long- and short-term user interests through hierarchical gating networks.
    \item \textbf{GRU4Rec}~\cite{hidasi2015session}: adopts a GRU-based recurrent network to encode the interaction sequence.
    \item \textbf{BERT4Rec}~\cite{sun2019bert4rec}: pretrains a bidirectional Transformer with a masked-item prediction objective.
    \item \textbf{FDSA}~\cite{zhang2019feature}: introduces a feature-level self-attention block that complements item-level attention.
    \item \textbf{SASRec}~\cite{kang2018self}: uses a unidirectional self-attention network for next-item prediction.
    \item \textbf{S$^3$-Rec}~\cite{zhou2020s3}: pretrains a sequential encoder with mutual-information-based self-supervision over items, attributes, and segments.
\end{itemize}

\noindent \textit{Generative recommenders:}
\begin{itemize}
    \item \textbf{TIGER}~\cite{rajput2023recommender}: represents each item by an RQ-VAE-based hierarchical semantic ID and generates the target ID autoregressively.
    \item \textbf{HSTU}~\cite{zhai2024actions}: reformulates recommendation as a generative modeling task over user actions and proposes the Hierarchical Sequential Transduction Unit, a linear-complexity architecture that scales generative recommendation to industrial settings.
    \item \textbf{COBRA}~\cite{yang2025sparse}: couples sparse semantic IDs with dense embeddings and retrieves candidates via approximate nearest neighbour search at inference time.
    \item \textbf{APAO-pointwise}~\cite{yu2026apao}: introduces a prefix-aware optimization objective to mitigate the training--inference mismatch of beam search. We adopt the pointwise variant rather than the pairwise one because the pairwise mode incurs a prohibitive inference cost, and the original paper of APAO also reports that its production deployment uses the pointwise mode.
    \item \textbf{ActionPiece}~\cite{hou2025actionpiece}: tokenizes user actions with a context-aware sub-action vocabulary that adapts the granularity of generation.
\end{itemize}

The baseline results on the Amazon datasets are taken from~\cite{yang2025sparse,hou2025actionpiece} under the same evaluation protocol, and the results of HSTU and ActionPiece are directly sourced from~\cite{hou2025actionpiece}.

\textit{Implementation details.}
All experiments are conducted on 2$\times$ NVIDIA H20 GPUs. We adopt a semantic ID depth of $L=4$ with a layer-wise decreasing codebook configuration $(|\mathcal{C}_1|, |\mathcal{C}_2|, |\mathcal{C}_3|, |\mathcal{C}_4|) = (512, 256, 128, 64)$ inspired by PLUM~\cite{he2025plum}. All four layers are learned semantic codebooks. The codebook configuration analysis is presented in Section~\ref{sec:codebook}. The encoder and each of the two DPD decoder towers share the same Transformer configuration with 2 layers and 4 attention heads. The embedding dimension is 128, the attention dimension is 512, and the FFN hidden size is 1024. The model is trained for 200 epochs at a batch size of 256, using the Adam optimizer with warmup. Early stopping is applied based on the validation performance, and all results are averaged over 3 random seeds. In addition to the full BARGE, we also report \textbf{BARGE-base}, a variant that keeps all model components but replaces our 4-layer learned codebook with TIGER's 3-layer codebook plus a random collision-resolving ID, so as to isolate the gain of the three structural modules from that of our codebook design.

\subsection{Overall Performance Comparison}

\begin{table*}[t]
\centering
\caption{Overall performance comparison on two Amazon datasets. Best result in bold; the strongest baseline is underlined.}
\label{tab:overall_amazon}
\begin{tabular}{>{\centering}p{1.9cm} p{2.56cm} | >{\centering}p{1.0cm} >{\centering}p{1.0cm} >{\centering}p{1.0cm} >{\centering}p{1.0cm} | >{\centering}p{1.0cm} >{\centering}p{1.0cm} >{\centering}p{1.0cm} >{\centering\arraybackslash}p{1.0cm}}
\toprule
& \multirow{2}{*}{Methods} & \multicolumn{4}{c}{Beauty} & \multicolumn{4}{c}{Sports and Outdoors} \\
\cmidrule(lr){3-6} \cmidrule(lr){7-10}
& & R@5 & N@5 & R@10 & N@10 & R@5 & N@5 & R@10 & N@10 \\
\midrule
\multirow{8}{*}{Non-Generative}
& P5 & 0.0163 & 0.0107 & 0.0254 & 0.0136 & 0.0061 & 0.0041 & 0.0095 & 0.0052 \\
& Caser & 0.0205 & 0.0131 & 0.0347 & 0.0176 & 0.0116 & 0.0072 & 0.0194 & 0.0097 \\
& HGN & 0.0325 & 0.0206 & 0.0512 & 0.0266 & 0.0189 & 0.0120 & 0.0313 & 0.0159 \\
& GRU4Rec & 0.0164 & 0.0099 & 0.0283 & 0.0137 & 0.0129 & 0.0086 & 0.0204 & 0.0110 \\
& BERT4Rec & 0.0203 & 0.0124 & 0.0347 & 0.0170 & 0.0115 & 0.0075 & 0.0191 & 0.0099 \\
& FDSA & 0.0267 & 0.0163 & 0.0407 & 0.0208 & 0.0182 & 0.0122 & 0.0288 & 0.0156 \\
& SASRec & 0.0337 & 0.0225 & 0.0536 & 0.0289 & 0.0177 & 0.0111 & 0.0309 & 0.0153 \\
& S$^3$-Rec & 0.0359 & 0.0218 & 0.0613 & 0.0299 & 0.0251 & 0.0161 & 0.0385 & 0.0204 \\
\midrule
\multirow{6}{*}{Generative}
& TIGER & 0.0454 & 0.0321 & 0.0648 & 0.0384 & 0.0264 & 0.0181 & 0.0400 & 0.0225 \\
& COBRA & \underline{0.0537} & \underline{0.0395} & 0.0725 & \underline{0.0456} & 0.0305 & \underline{0.0215} & 0.0434 & 0.0257 \\
& HSTU & 0.0469 & 0.0314 & 0.0704 & 0.0389 & 0.0258 & 0.0165 & 0.0414 & 0.0215 \\
& ActionPiece & 0.0511 & 0.0340 & 0.0775 & 0.0424 & \underline{0.0316} & 0.0205 & \underline{0.0500} & \underline{0.0264} \\
& APAO-pointwise & 0.0530 & 0.0368 & \underline{0.0795} & 0.0453 & 0.0283 & 0.0186 & 0.0444 & 0.0237 \\
& BARGE-base & 0.0598 & 0.0420 & 0.0896 & 0.0515 & 0.0337 & 0.0229 & 0.0513 & 0.0285 \\
& BARGE & \textbf{0.0654} & \textbf{0.0460} & \textbf{0.0927} & \textbf{0.0547} & \textbf{0.0369} & \textbf{0.0252} & \textbf{0.0544} & \textbf{0.0308} \\
\bottomrule
\end{tabular}
\end{table*}

Table~\ref{tab:overall_amazon} reports the overall performance of BARGE against discriminative and generative baselines on the two datasets. We summarize the key observations below:

\noindent$\bullet$ \textbf{Overall superiority of BARGE.} BARGE achieves the best score on every metric across both datasets, with representative gains of $+16.6\%$/$+8.8\%$ R@10 on Beauty/Sports over the strongest baseline. The gains are uniform over $K$ and over both recall and ranking metrics, indicating a structural improvement rather than a mere shift of operating point.

\noindent$\bullet$ \textbf{Advantage over discriminative recommenders.} Discriminative baselines trail the generative family by a clear margin, and BARGE further enlarges this gap. Representing each item with a single ID embedding forces items with related but non-identical semantics to be aligned through interaction signals alone, which is fragile under data sparsity. BARGE instead encodes each item as a hierarchical semantic ID and predicts it layer by layer, so coarse-grained aspects such as category and brand are shared across related items and only the fine-grained distinctions need to be learned from behavior. This yields stronger few-shot generalization on cold and tail items.

\noindent$\bullet$ \textbf{Advantage over generative recommenders.} Compared with prior generative recommenders, BARGE consistently ranks first, and we attribute the improvement to a more structured treatment of the hierarchical semantic ID on both the encoder and the decoder sides. Existing methods are trained purely under a per-token likelihood objective, so shallow-layer errors are not penalised beyond their own step and tend to cascade into deeper layers. In addition, they commit to a single deterministic decoding path, which makes off-path items effectively unreachable no matter how sharp the token-level distribution is. Remedies such as attaching an extra ANN retrieval stage or refining the training loss only address part of the picture, and often reintroduce the full-catalogue cost and other overheads that a purely generative paradigm was meant to remove. BARGE tackles these two weaknesses jointly through three complementary modules.

\noindent$\bullet$ \textbf{Disentangling the codebook from the model design.} BARGE-base already surpasses every prior generative baseline on both datasets, isolating the contribution of ICA, HPR, and DPD from that of our codebook design. The remaining gap to the full BARGE shows that the 4-layer codebook in Section~\ref{sec:codebook} is complementary to the structural modules rather than a substitute for them.

\begin{table}[h]
\centering
\caption{Performance comparison on the Tencent commercial media platform offline test.}
\label{tab:novel}
\setlength{\tabcolsep}{5pt}
\begin{tabular}{p{2.3cm} >{\centering}p{1.1cm} >{\centering}p{1.1cm} >{\centering}p{1.1cm} >{\centering\arraybackslash}p{1.1cm}}
\toprule
& Hit@5 & Hit@10 & Hit@20 & Hit@50 \\
\midrule
GraphSAGE-based & 0.2932 & 0.3743 & 0.4650 & 0.5951 \\
NANN  & 0.4416 & 0.4946 & 0.5636 & 0.6760 \\
OneRec      & \underline{0.5459} & \underline{0.6132} & \underline{0.6729} & \underline{0.7348} \\
BARGE       & \textbf{0.6015} & \textbf{0.6510} & \textbf{0.6967} & \textbf{0.7520} \\
\bottomrule
\end{tabular}
\end{table}

On the Tencent commercial media platform offline test (Table~\ref{tab:novel}), BARGE consistently outperforms all baselines. GraphSAGE-based~\cite{hamilton2017inductive} and NANN~\cite{chen2022approximate} both operate on atomic item embeddings and cannot express the shared coarse-grained structure among semantically related items, which is especially costly on a catalogue with hundreds of thousands of long-tail items. OneRec narrows this gap through a generative formulation over semantic IDs, but its single-path token-level objective still leaves shallow-layer errors and off-path items as the dominant failure modes. BARGE keeps the generative backbone and repairs these two weaknesses, with the margin most pronounced at the top of the ranked list. This advantage also carries over to a catalogue orders of magnitude larger than the academic benchmarks. Such scalability suggests that path-level supervision and dual-path decoding grow in importance as the codebook has to cover a more heterogeneous item population in real-world recommendation scenarios.

\begin{table}[h]
\centering
\caption{Efficiency comparison between TIGER and BARGE.}
\label{tab:efficiency}
\begin{tabular}{l >{\centering}p{1.7cm} >{\centering}p{1.8cm} >{\centering\arraybackslash}p{1.8cm}}
\toprule
& Params & Train (s/epoch) & Infer (s/epoch) \\
\midrule
TIGER & 22.71\,M & 22 & 17 \\
BARGE & 19.91\,M & 24 & 18 \\
\bottomrule
\end{tabular}
\end{table}

\subsection{Efficiency Analysis}
\label{sec:efficiency}

To assess whether ICA, HPR, and DPD introduce non-trivial overhead, we compare BARGE against TIGER on Amazon Beauty under the same training and inference configuration. As shown in Table~\ref{tab:efficiency}, although BARGE adds three structural modules on top of the backbone, its total parameter count is comparable to TIGER. This is because we use a 2-layer encoder rather than the 4-layer encoder in TIGER: the saving on the encoder side absorbs the extra parameters of ICA, the HPR scorers, and the DPD dual-decoder, so that the introduced structural changes do not translate into a noticeable increase in model size. The added per-epoch cost is also modest, since the two DPD towers share the same encoder and run in parallel. Overall, BARGE delivers its accuracy gains without inflating model size or wall-clock cost.

\begin{table*}[t]
\centering
\caption{Ablation study on the two Amazon datasets.}
\label{tab:ablation}
\begin{tabular}{>{\centering}p{2.4cm} p{2.56cm} | >{\centering}p{1.0cm} >{\centering}p{1.0cm} >{\centering}p{1.0cm} >{\centering}p{1.0cm} | >{\centering}p{1.0cm} >{\centering}p{1.0cm} >{\centering}p{1.0cm} >{\centering\arraybackslash}p{1.0cm}}
\toprule
& \multirow{2}{*}{Variant} & \multicolumn{4}{c}{Beauty} & \multicolumn{4}{c}{Sports and Outdoors} \\
\cmidrule(lr){3-6} \cmidrule(lr){7-10}
& & R@5 & N@5 & R@10 & N@10 & R@5 & N@5 & R@10 & N@10 \\
\midrule
\multirow{4}{*}{Component}
& BARGE (full) & \textbf{0.0654} & \textbf{0.0460} & 0.0927 & \textbf{0.0547} & \textbf{0.0369} & \textbf{0.0252} & \textbf{0.0544} & \textbf{0.0308} \\
& BARGE w/ ICA & 0.0577 & 0.0393 & 0.0859 & 0.0483 & 0.0276 & 0.0181 & 0.0446 & 0.0235 \\
& BARGE w/ HPR & 0.0592 & 0.0413 & 0.0864 & 0.0500 & 0.0304 & 0.0202 & 0.0481 & 0.0258 \\
& BARGE w/ DPD & 0.0629 & 0.0437 & 0.0913 & 0.0529 & 0.0350 & 0.0237 & 0.0527 & 0.0294 \\
\midrule
\multirow{4}{*}{\shortstack{OR-fusion\\function}}
& LSE  & \textbf{0.0654} & \textbf{0.0460} & 0.0927 & \textbf{0.0547} & \textbf{0.0369} & \textbf{0.0252} & \textbf{0.0544} & \textbf{0.0308} \\
& Max  & 0.0634 & 0.0451 & 0.0913 & 0.0540 & 0.0351 & 0.0239 & 0.0528 & 0.0296 \\
& Mean & 0.0621 & 0.0436 & 0.0888 & 0.0522 & 0.0341 & 0.0229 & 0.0526 & 0.0288 \\
& RRF  & \textbf{0.0654} & 0.0451 & \textbf{0.0931} & 0.0540 & 0.0360 & 0.0245 & 0.0538 & 0.0303 \\
\midrule
\multirow{1}{*}{Rotation}
& Random $R$ (frozen) & 0.0591 & 0.0420 & 0.0845 & 0.0502 & 0.0309 & 0.0213 & 0.0472 & 0.0265 \\
\bottomrule
\end{tabular}
\end{table*}

\subsection{Ablation Study}
\label{sec:ablation}

We conduct ablation studies along three complementary axes, all reported in Table~\ref{tab:ablation}. The first axis (\textit{Component}) progressively adds one of the three modules, in order to isolate the individual contribution of ICA, HPR, and DPD. The second axis (\textit{OR-fusion function}) fixes the full BARGE backbone and varies the fusion operator $f(\cdot,\cdot)$ in Eq.~\eqref{eq:dpd_fusion}, in order to verify that the OR semantics, rather than a particular implementation of it, is what drives the gain of DPD. We compare four representative instantiations: (1) \textbf{LSE}, $f{=}\log(\exp s^{(A)}{+}\exp s^{(B)})$, a soft-OR that smoothly favors the channel with the higher score; (2) \textbf{Max}, $f{=}\max(s^{(A)},s^{(B)})$, which most directly implements the OR semantics by trusting whichever channel ranks the item higher; (3) \textbf{Mean}, $f{=}\tfrac{1}{2}(s^{(A)}{+}s^{(B)})$, an AND-style average that requires both channels to agree; and (4) \textbf{RRF}, the reciprocal-rank fusion $f{=}\tfrac{1}{k+\mathrm{rank}^{(A)}(v)}+\tfrac{1}{k+\mathrm{rank}^{(B)}(v)}$, which is rank-based. The third axis (\textit{Rotation}) replaces the learned orthogonal rotation $R$ in DPD with a fixed random orthogonal matrix, in order to verify that the gain of DPD comes from \emph{learning} a task-specific orthogonal decomposition rather than merely from any orthogonal split.

\noindent$\bullet$ \textbf{Component-level results.} All three single-module variants outperform TIGER, confirming that each module delivers gains independently. Among them, BARGE w/ DPD attains the strongest single-module performance, as decoupling an item into two orthogonal channels directly enlarges the recoverable semantic coverage. BARGE w/ HPR ranks second, showing that path-level reranking effectively suppresses intra-path drift even within a single channel, while BARGE w/ ICA confirms the value of restoring item-level structure at the encoder. The full BARGE further surpasses all single-module variants, demonstrating the complementarity of the three modules: ICA, HPR, and DPD address non-overlapping failure sources at the encoder, the path level, and across channels, respectively.

\noindent$\bullet$ \textbf{OR-fusion function results.} LSE achieves strong results on most metrics across the two datasets, validating the soft-OR formulation as the default choice for $f$. The three OR-style operators (LSE, Max, RRF) all clearly outperform the AND-style Mean, corroborating the design principle of DPD: an item should be recovered as long as at least one channel ranks it highly. LSE consistently dominates Max because Max relies on a hard $\arg\max$ that is sensitive to score-scale differences between channels, whereas LSE smoothly aggregates both sides and remains robust when the two scores are close. RRF is competitive with LSE on Beauty but trails on Sports, suggesting that score-based soft-OR carries finer-grained signal than rank-based fusion when channel scores are well-calibrated by the per-channel HPR.

\noindent$\bullet$ \textbf{Rotation results.} Replacing the learned $R$ with a frozen random orthogonal matrix consistently degrades performance on both datasets, since a random $R$ splits the embedding space along directions unrelated to the recommendation objective. This confirms that the gain of DPD stems from \emph{learning} a task-aware orthogonal decomposition, not from any arbitrary orthogonal split.

\subsection{Hyperparameter Sensitivity Analysis}

We study the sensitivity of two key HPR hyperparameters: the reranking weight $\lambda$ and the reranker Top-N.

\begin{figure}[t]
    \centering
    \includegraphics[width=\columnwidth]{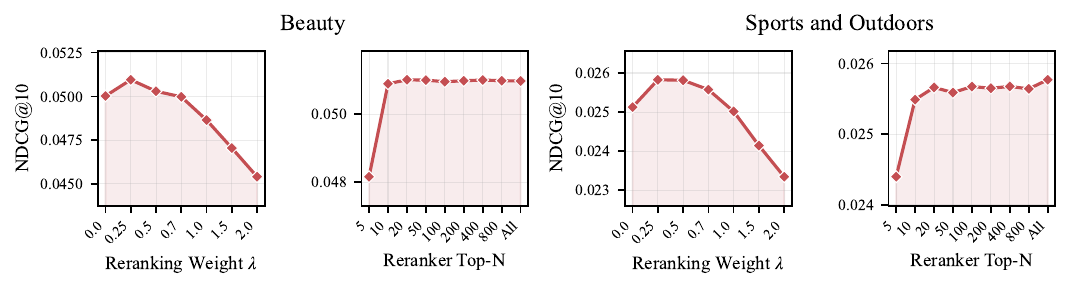}
    \caption{\small Sensitivity of HPR hyperparameters (NDCG@10).}
    \label{fig:hpr_sensitivity}
\end{figure}

\textit{Reranking weight $\lambda$.} As shown in Fig.~\ref{fig:hpr_sensitivity} (left two panels), we vary $\lambda \in \{0.0, 0.25, 0.5, 0.7, 1.0, 1.5, 2.0\}$. The performance exhibits an inverted-U pattern: a moderate $\lambda$ balances the generation likelihood (which captures local token-level fluency) and the reranking signal (which captures global path-level semantic coherence), while an excessively large $\lambda$ causes the reranker to dominate and override valid generation probabilities. We select $\lambda = 0.25$.

\textit{Reranker Top-N.} As shown in Fig.~\ref{fig:hpr_sensitivity} (right two panels), the performance improves rapidly as Top-N increases from small values, and then plateaus beyond approximately 400. This is expected: once the scoring pool is large enough to include the correct path with high probability, further expansion introduces only low-scoring candidates that do not affect the final ranking. We select Top-N $= 400$.

\subsection{Analysis of Hierarchical Semantic Drift}
\label{sec:motivation}

To understand how prefix errors propagate across decoding layers, we compare the prediction quality on the TIGER baseline under two modes: Teacher Forcing (TF), which feeds the ground-truth prefix to the decoder, and Autoregressive (AR), which uses the model-selected prefix as in standard inference. The experiments are conducted on the Beauty test set with greedy decoding. The TF--AR gap directly measures the damage caused by prefix errors.

As shown in Table~\ref{tab:drift_tf_ar}, the deepest layer $c_3$ achieves a target probability of $0.787$ under TF but collapses to $0.015$ under AR, a $52\times$ drop driven entirely by the propagation of prefix errors. The bottom block further isolates the cascade mechanism: at $c_3$, the per-layer accuracy is $77.0\%$ when the AR prefix is entirely correct but drops to $0.6\%$ when prefix errors exist, yielding a $128\times$ gap that is dominated by prefix correctness rather than by the capacity of the layer itself. While beam search ($B>1$) partially mitigates the issue by maintaining multiple candidate paths, it does not address the root cause: without any mechanism to assess cross-layer semantic coherence, erroneous paths continue to propagate, and enlarging $B$ is impractical in production because inference latency scales linearly with $B$ while GPU memory grows as $B\times L$. These observations motivate HPR as a principled reranking mechanism that corrects semantically inconsistent paths without relying on brute-force beam expansion.

\begin{table}[H]
\centering
\caption{Per-layer prediction quality under Teacher Forcing (TF) and Autoregressive (AR) modes on Amazon Beauty. The last two rows split the AR results at $c_3$ by whether the autoregressive prefix is entirely correct.}
\label{tab:drift_tf_ar}
\begin{tabular}{p{2.6cm} >{\centering}p{1.2cm} >{\centering}p{1.4cm} >{\centering\arraybackslash}p{1.4cm}}
\toprule
\textbf{Layer / Mode} & \textbf{Mismatch$\downarrow$} & \textbf{Rank$\downarrow$} & \textbf{Prob$\uparrow$} \\
\midrule
$c_1$ (TF / AR)            & 0.915 & 66.9  & 0.046 \\
\midrule
$c_2$ (TF)                 & 0.864 & 25.3  & 0.102 \\
$c_2$ (AR)                 & 0.981 & 108.2 & 0.015 \\
\midrule
$c_3$ (TF)                 & 0.195 & 7.4   & 0.787 \\
$c_3$ (AR)                 & 0.984 & 119.5 & 0.015 \\
\midrule
$c_3$ (AR, prefix correct) & 0.230 & 3.7   & 0.738 \\
$c_3$ (AR, prefix error)   & 0.994 & 121.1 & 0.006 \\
\bottomrule
\end{tabular}
\end{table}

\subsection{DPD Analysis}
\label{sec:ablation_dpd}

We verify two properties of DPD that justify its design without relying on idealised assumptions: (i) the learnable rotation $R$ is used in a non-trivial way, and (ii) the two channels propose empirically complementary top-$K$ candidate sets, so that a non-trivial fraction of OR-fusion hits is contributed exclusively by a single channel.

\textit{Diagnostic of the learned rotation $R$.}
We probe the trained $R$ on Beauty and Sports along two axes: a structural test (how exactly orthogonality holds and how far $R$ deviates from $I_D$) and a functional test (replacing $R$ with $I_D$ at inference time), as summarized in Table~\ref{tab:rotation_diag}.

\begin{table}[h]
\centering
\caption{Diagnostic of the learned rotation $R$ ($D{=}32$).}
\label{tab:rotation_diag}
\begin{tabular}{p{3.2cm} >{\centering}p{1.6cm} >{\centering\arraybackslash}p{1.6cm}}
\toprule
& Beauty & Sports \\
\midrule
$\|R-I\|_F/\sqrt{D}$        & 1.361 & 1.243 \\
$\|R^{\top}R-I\|_F$         & 9.6\text{e-}6 & 9.0\text{e-}6 \\
Recon loss with $R{:=}I$    & 0.2067 & 0.2500 \\
Recon loss with $R$         & 0.1621 & 0.2012 \\
$\Delta$Recon loss          & $-0.0445$ & $-0.0488$ \\
\bottomrule
\end{tabular}
\end{table}

The Householder parameterisation enforces $R^{\top}R\!=\!I_D$ to numerical precision ($\sim\!10^{-6}$), while $\|R-I\|_F/\sqrt{D}\!\approx\!1.2$--$1.4$ confirms that $R$ moves substantially away from the identity. The trained $R$ consistently reduces the reconstruction loss by $0.04$--$0.05$ relative to the $R{:=}I_D$ baseline on both datasets, showing that $R$ encodes information actively used by the OSQ-VAE quantizer rather than collapsing to a trivial solution.

\textit{Empirical complementarity of the two channels.}
We next measure on the Amazon test sets whether the two channel-private decoders make complementary mistakes, which is the empirical property that OR-fusion relies on. For every test sample we record (i) whether each channel hits the ground-truth item at $K{=}10$ and (ii) each channel's top-$K$ candidate pool $V_K^{(c)}$. From these we report in Table~\ref{tab:dpd_corr} the marginal hit rates of the two channels and of OR-fusion, the Jaccard similarity $|V_K^{(A)}\cap V_K^{(B)}|/|V_K^{(A)}\cup V_K^{(B)}|$, and, among the test samples hit by OR-fusion, the fraction whose ground-truth item is present in both channels' top-$K$ pools versus exclusively in one.

\begin{table}[h]
\centering
\caption{Empirical complementarity of the two DPD channels on the Amazon test sets at $K{=}10$. The last three rows sum to $100\%$ within each column.}
\label{tab:dpd_corr}
\begin{tabular}{p{3.7cm} >{\centering}p{1.3cm} >{\centering\arraybackslash}p{1.3cm}}
\toprule
& Beauty & Sports \\
\midrule
Hit rate, view\,A           & 0.0879 & 0.0499 \\
Hit rate, view\,B           & 0.0873 & 0.0496 \\
Hit rate, OR-fusion         & \textbf{0.0928} & \textbf{0.0544} \\
\midrule
Jaccard$(V^{(A)},V^{(B)})$  & 0.183 & 0.172 \\
\midrule
GT $\in V^{(A)}\cap V^{(B)}$ (Shared)  & 84.4\% & 75.7\% \\
GT $\in V^{(A)}\setminus V^{(B)}$ (Excl.\,A) & 8.3\% & 11.8\% \\
GT $\in V^{(B)}\setminus V^{(A)}$ (Excl.\,B) & 7.3\% & 12.6\% \\
\bottomrule
\end{tabular}
\end{table}

Three consistent findings emerge across both datasets. First, OR-fusion strictly improves the hit rate over either channel alone, and these gains directly translate into the improvements reported in Table~\ref{tab:overall_amazon}. Second, the two top-$K$ pools are highly complementary, with an average Jaccard of only $0.18$ on Beauty and $0.17$ on Sports. Third, a non-trivial fraction of OR-fusion hits comes from ground-truth items reachable through only one channel, confirming that the two channels rescue genuinely different ground-truth items rather than redundantly agreeing on easy ones. Together, these measurements support the OR-fusion design empirically: as long as the channels do not collapse to identical failure patterns, OR-fusion yields a measurable gain.

\begin{figure}[H]
    \centering
    \includegraphics[width=\columnwidth]{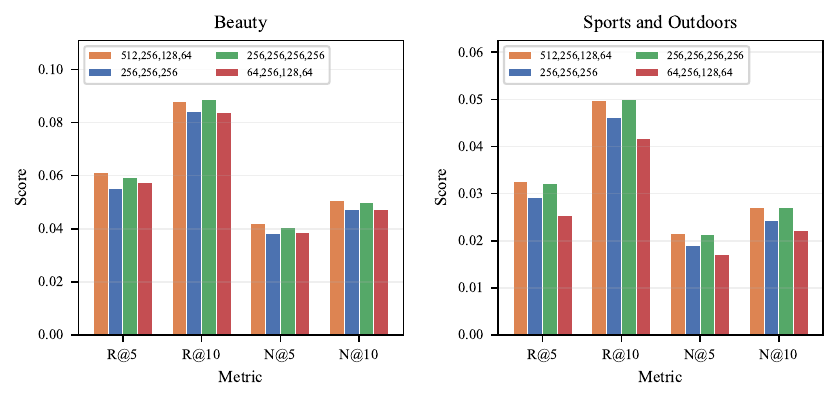}
    \caption{\small Codebook configuration analysis on Beauty and Sports. Four configurations are compared across R@5, N@5, R@10, and N@10.}
    \label{fig:codebook_analysis}
\end{figure}

\subsection{Codebook Configuration Analysis}
\label{sec:codebook}

BARGE adopts a fully learned 4-layer codebook ($L{=}4$) with layer-wise decreasing sizes $(|\mathcal{C}_1|,\dots,|\mathcal{C}_4|)\!=\!(512, 256, 128, 64)$, which differs from the TIGER convention~\cite{rajput2023recommender} of a 3-layer uniform codebook plus a collision-resolving ID. Two design choices warrant explanation.

\textit{Why no random collision-resolving ID at the last layer.}
Item-level structural modelling requires \emph{every} semantic-ID layer, including the last, to carry genuine semantic information; appending a random ID at the deepest layer would inject a non-semantic bit into the very position where ICA and HPR operate, breaking the layer-wise structural signal that these modules rely on. To assess whether this concession is even necessary, we run a controlled comparison on TIGER between its original 3-layer codebook with a random collision-resolving ID and a variant in which the random bit is replaced by an additional fully-learned semantic layer of the same size: the two settings differ by less than 0.1\% on all recommendation metrics, indicating that whatever benefit the random ID brings can be matched by simply turning that slot into a semantic layer.

\textit{Why layer-wise decreasing sizes.}
Following the progressive layer-wise modelling principle of PLUM~\cite{he2025plum}, we let the first layer use a larger codebook to cover diverse coarse-grained categories and shrink subsequent layers as they refine the representation within increasingly constrained semantic subtrees. Beyond the conceptual fit, this configuration is also cheaper: compared with a 4-layer uniform $(256\!\times\!4)$ codebook, our $(512, 256, 128, 64)$ schedule reduces the total number of codeword embeddings from $1024$ to $960$, and the per-layer output-head parameters scale accordingly.

To support these two choices empirically, we compare four configurations: (1) $(512, 256, 128, 64)$, our final setting; (2) $(256, 256, 256)$ plus a random collision-resolving ID, which is the original TIGER configuration; (3) $(256, 256, 256, 256)$, a 4-layer uniform configuration with a fully-learned last layer; and (4) $(64, 256, 128, 64)$, a variant with a deliberately small first-layer codebook. Fig.~\ref{fig:codebook_analysis} yields two observations consistent with the two design choices above:

\textit{(1) Depth matters, and layer-wise decreasing is the most effective allocation.} The 4-layer uniform consistently outperforms the TIGER-style 3-layer codebook with a collision-resolving ID on both datasets, confirming that an additional semantic layer is the right way to expand the ID space. On top of this, our $(512, 256, 128, 64)$ schedule further outperforms the 4-layer uniform baseline while using fewer parameters, validating the PLUM-style decreasing allocation.

\textit{(2) Capacity should concentrate at coarse-grained layers.} Comparing $(512, 256, 128, 64)$ with $(64, 256, 128, 64)$, shrinking only the first layer from 512 to 64 causes consistent degradation across all metrics on both datasets. This confirms that the first layer, which defines the coarse-grained semantic partitioning, requires sufficient capacity to establish well-separated category boundaries, exactly the regime in which a larger first-layer codebook pays off most.

\subsection{Qualitative Analysis}
\label{sec:qualitative}

\begin{figure}[t]
    \centering
    \includegraphics[width=\columnwidth]{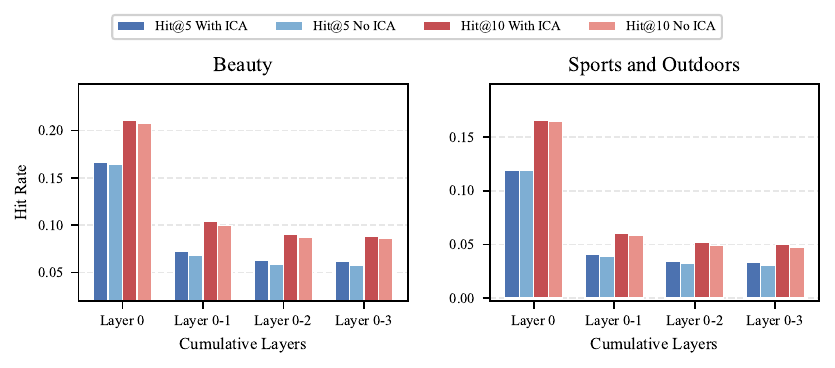}
    \caption{\small Cumulative hit rate comparison with and without ICA across semantic ID layers. The advantage of ICA grows as layers accumulate, demonstrating stronger resistance to semantic drift at deeper layers.}
    \label{fig:ica_hit_rate}
\end{figure}

\textit{ICA: Per-layer cumulative hit rate.}
Fig.~\ref{fig:ica_hit_rate} compares the cumulative hit rate with and without ICA across semantic ID layers. The advantage of ICA becomes more pronounced as layers accumulate: at $c_1$ the gap is relatively small, but by $c_4$ the ICA-equipped model maintains a clearly higher hit rate. This widening gap demonstrates that ICA effectively mitigates semantic drift at deeper layers by injecting item-level context that provides the discriminative signal needed to keep the decoding path on track.

\begin{figure}[t]
    \centering
    \includegraphics[width=\columnwidth]{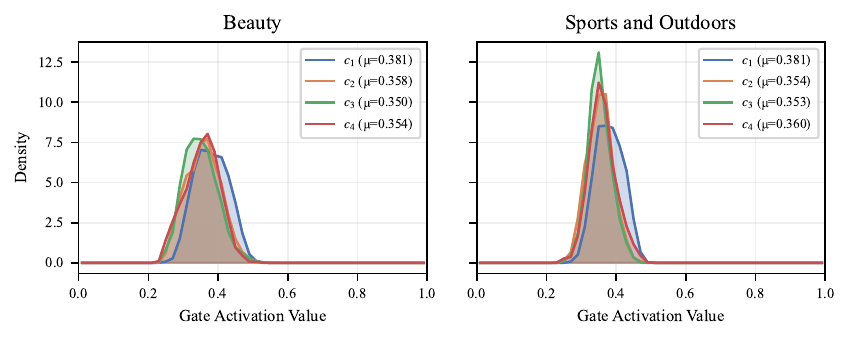}
    \caption{\small Gate activation value distribution across semantic ID layers. Gate values are stably concentrated around 0.35--0.38, indicating selective and consistent injection of item-level context.}
    \label{fig:ica_gate}
\end{figure}

\textit{ICA: Gate activation distribution.}
Fig.~\ref{fig:ica_gate} visualises the distribution of the gate activations across all four semantic ID layers. The gate values are concentrated around 0.35--0.38 on both datasets, which indicates that the gating network selectively injects item-level context rather than overwriting token representations. The distributions remain consistent across layers ($c_1$ through $c_4$), which suggests that item-level context is stably beneficial at all levels of the hierarchy.

\textit{HPR: Drift-recovery case study.}
To examine when HPR is most beneficial, we sweep the entire Sports test set and identify samples for which vanilla beam search drops the ground-truth (GT) path at some intermediate layer (i.e., its NTP-only beam rank exceeds the beam width $K{=}20$), while HPR rescues it back into the surviving beam by fusing the per-layer dual-tower path score. Two representative cases are presented below, which illustrate two drift patterns.

\noindent\textit{Case~A: Single-layer drift on a hunting/tactical sequence.}
\begin{caseInput}
\footnotesize
A user's last five interactions form a coherent hunting and tactical chain:
\begin{enumerate}[leftmargin=1.4em,itemsep=0pt,topsep=2pt]
  \item NcStar -- Paintball \& Airsoft / Paintball Sights
  \item Plano -- Hunting \& Fishing / Tackle Boxes
  \item Mtech USA -- Hunting \& Tactical Knives / Hunting Knives
  \item (Unbranded) -- Hunting Accessories / Cleaning \& Maintenance Products
  \item Predator -- Hunting Accessories / Targets \& Accessories
\end{enumerate}
\end{caseInput}
\begin{caseGT}
\footnotesize
\textbf{Sports \& Outdoors} (generic, no fine-grained sub-category)\\[1pt]
\textit{Semantic ID:} $(243,\,182,\,45,\,41)$
\end{caseGT}
\begin{caseDrift}
\footnotesize
\centering
\setlength{\tabcolsep}{6pt}
\renewcommand{\arraystretch}{1.05}
\begin{tabular}{c c c c}
\textbf{Layer} & \textbf{NTP rank} & \textbf{HPR rank} & \textbf{$\Delta$} \\
\hline
$c_2$ & 33 (dropped) & \textbf{14} & $+19$ \\
$c_3$ & 22 (dropped) & \textbf{13} & $+9$
\end{tabular}
\end{caseDrift}
\noindent After a long hunting-themed history, vanilla NTP overcommits to fine-grained hunting sub-categories at $c_2$ and pushes the more generic GT outside the top-20. The path-level score of HPR rewards prefixes that remain compatible with the broader category, which recovers the GT to rank~14.

\smallskip
\noindent\textit{Case~B: Multi-layer drift on a mixed-interest sequence.}
\begin{caseInput}
\footnotesize
The history mixes cycling, swimming and tactical interests:
\begin{enumerate}[leftmargin=1.4em,itemsep=0pt,topsep=2pt]
  \item Shimano -- Cycling / Drivetrain Components / Chains
  \item (Unbranded) -- Boating \& Water Sports / Swimming Training Paddles
  \item Profile Design -- Cycling / Handlebars
  \item (Unbranded) -- Swimming / Men's Swimwear / Jammers
  \item MTECH USA XTREME -- Hunting \& Fishing / Tactical Knives
\end{enumerate}
\end{caseInput}
\begin{caseGT}
\footnotesize
\textbf{Aquatic Fitness Equipment} (Exercise \& Fitness / Accessories)\\[1pt]
\textit{Semantic ID:} $(118,\,158,\,30,\,15)$
\end{caseGT}
\begin{caseDrift}
\footnotesize
\centering
\setlength{\tabcolsep}{6pt}
\renewcommand{\arraystretch}{1.05}
\begin{tabular}{c c c c}
\textbf{Layer} & \textbf{NTP rank} & \textbf{HPR rank} & \textbf{$\Delta$} \\
\hline
$c_2$ & 32 (dropped) & \textbf{19} & $+13$ \\
$c_3$ & 23 (dropped) & \textbf{17} & $+6$ \\
$c_4$ & 23 (dropped) & \textbf{20} & $+3$
\end{tabular}
\end{caseDrift}
\noindent NTP-only decoding consistently drops the ground-truth path at every one of the three deep layers, whereas HPR successfully rescues it at all three layers simultaneously. This confirms that the per-layer reranker repeatedly intervenes whenever the locally most likely prefix misaligns with the global semantic plausibility implied by the full history of the user, effectively correcting beam search errors before they propagate further.

\smallskip
\noindent\textit{Summary.} Across the full test set, we observe such drift-recovery events on a non-trivial fraction of samples, with rank improvements ranging from a few positions to more than $30$. This qualitative picture is consistent with the aggregate gains reported in Table~\ref{tab:overall_amazon} and the cascade-drift statistics in Table~\ref{tab:drift_tf_ar}.

\subsection{Online A/B Test}
\label{sec:online_ab}

We further conducted an online A/B test on a Tencent commercial media platform, allocating 6\% of the live traffic to BARGE and comparing it against the incumbent multi-stage system. BARGE achieved statistically significant improvements on core engagement metrics, with click-through rate +0.60\%, click unique visitors +1.34\%, and total reading time +1.70\%, confirming that its offline gains translate into tangible business value in industrial-scale recommendation.

\section{Conclusion}

In this work, we revisit generative recommendation from a structural standpoint and ask where, along the encode-decode pipeline, fidelity to the recommendation task is actually lost. Our answer locates two specific failure points: token flattening dissolves item boundaries inside the encoder, and hierarchical autoregressive decoding accumulates drift across codebook layers. Around this diagnosis we build BARGE as three mutually orthogonal modules. ICA reinstates item-level granularity through cross-attention pooling with gated fusion, HPR rescores beams via per-layer dual-tower contrastive evaluation, and DPD decomposes item embeddings into orthogonal subspaces via OSQ-VAE and decodes them through paired branches over a shared encoder, each with channel-private HPR. Empirically, the three modules act on disjoint error sources and their gains add up rather than overlap, as evidenced on public benchmarks and the large-scale offline test on Tencent's commercial media platform. Beyond BARGE itself, we plan to further explore the use of hierarchical semantics in designing more powerful structures for generative recommendation.


\newpage

\bibliographystyle{IEEEtran}
\bibliography{IEEEabrv,IEEEexample}

\end{document}